# MODELING VENUS-LIKE WORLDS THROUGH TIME

Michael J. Way[1,2], Anthony Del Genio[1] and David S. Amundsen[1,3]
*[1]NASA Goddard Institute for Space Studies, New York, NY, USA*
*[2]Department of Physics & Astronomy, Uppsala University, Uppsala, Sweden,*
*[3]Department of Applied Physics & Applied Mathematics, Columbia University, NY, USA*

**Abstract**
We explore the atmospheric and surface history of a hypothetical paleo-Venus climate using a 3-D General Circulation Model. We constrain our model with the in-situ and remote sensing Venus data available today. Given that Venus and Earth are believed to be similar geochemically some aspects of Earth's history are also utilized. We demonstrate that it is possible for ancient Venus and Venus-like exoplanetary worlds to exist within the liquid water habitable zone with insolations up to nearly 2 times that of modern Earth.

**Introduction**
In a recent paper Way et al. (2016) (hereafter Paper I) demonstrated that the climatic history of Venus may have allowed for surface liquid water to exist for several billion years. Paper I utilized ROCKED-3D a planetary three-dimensional (3-D) General Circulation Model (Way et al. 2017). A number of assumptions were made in Paper I including: the use of a solar spectrum at 2.9Gya and 0.715Gya epochs, orbital parameters that remained unchanged from those of today, roughly modern Earth atmospheric gas constituents (no aerosols or anthropogenic gases were included, only 1bar N2, 400ppm $CO_2$, and 1ppm CH4) and pressures (1013mb), a shallow ocean whose volume was consistent with Deuterium to Hydrogen ratio (D/H) observations, and topographies consistent with those of modern Venus and a single simulation using modern Earth with a shallow 310m deep fully coupled dynamic ocean. Except for one simulation that used a faster rotation rate (16 times the length of modern Earth's sidereal day) all of the simulations were able to maintain surface liquid water. In the meantime, we have inserted a more accurate radiation scheme called SOCRATES (Edwards & Slingo 1996) that allows us to explore insolations up to 1.9 times that received by modern Earth. We have also explored a variety of boundary conditions involving different dynamic ocean depths as well as orbital configurations that include a tidally locked world as may have existed or may exist for similar exoplanetary worlds. In most of these additional cases we find that the planet is able to maintain liquid surface water while keeping the stratosphere relatively dry, hence avoiding the possibility of a moist greenhouse.

**Methods**
We utilized ROCKE-3D (Way et al. 2017) for the simulations in this study. ROCKE-3D is a 3-D GCM whose parent model ModelE2 (Schmidt et al. 2014) is used for Earth Climate studies. ROCKE-3D is an extension of ModelE2 that allows one to explore non-Earth specific worlds with differing atmospheric constituents, pressures, orbital parameters, gravities, incident stellar flux and other parameters that may be different on terrestrial worlds within our solar system and exoplanetary worlds outside. The choice of which model boundary conditions to assume for a paleo-Venus type world are constrained by what little data we have from ground, space and in-situ observations. The Venus topography used herein is taken from Magellan satellite data as described in Paper I. The value of the lower bound on the D/H ratio (Donahue & Russell 1997) is such that it is possible to make a Dune-like land planet with as little as 10m water equivalent layer. We called 4 such simulations "Land Planets" (L1, L2, L1M and L2M, where M=Modern Solar Spectrum/insolation). In these simulations water is deposited as lakes at model start at the lowest elevation grid points. These are then allowed to expand, contract or form elsewhere as the model evolves and the competition between precipitation, evaporation and run-off take place (see Way & Wang 2017). At the same time given the uncertainties in the history of the topography of Venus it is perhaps equally possible that it could have been a shallow aquaplanet in its early history. Hence our aquaplanet simulations utilize a shallow aquaplanet configuration of depth 158m (A1, A2, A1M and A2M).

The choice of 158 meters depth (specifically it is the lower boundary of one of our dynamic ocean levels) was a compromise between the 310 meter depth we have used for the 4 simulations with the same land/sea mask as used in Paper I (V1, V2, V1M and V2M), and the 100 meter depth typically used in Qflux=0 exoplanet simulations. We also explore the possibility that ancient Venus may have experienced a tidally locked, synchronously rotating phase. Previous work by Dobrovolskis & Ingersoll (1980), and Corriea & Laskar (2003) and more recent work by Barnes (2017) have explored the possibility that a thinner 1-bar type atmosphere would not cause the same tidal torqueing effects as the present day ~90-bar atmosphere that is theorized to explain Venus' retrograde rotation. However, see Leconte et al. (2015) for an opposing viewpoint. Thus, we explore what the climate of ancient Venus may have been like if it were tidally locked at one epoch of its history. This may be particularly relevant to exoplanetary studies since according to Barnes (2017) it is possible that any planet around a G-star including and interior to the orbit of Earth may eventually reach a tidally locked state. This type of system has rarely been explored and only for a subset of hypothetical worlds (e.g. Yang et al. 2014).

**Table 1: Experiments and resulting temperatures:** Temperature values for non-tidally locked worlds are from 1/6 of a diurnal cycle, whereas the tidally locked numbers come from a 100 orbit average.

| Experiment | Land/Ocean | Spin Rate | Stellar Spectrum | Insolation | Temp Mean | Temp Max | Temp Min |
|---|---|---|---|---|---|---|---|
| L1 (001L) | Venus 10m | Present Day | 2.9Gya | 1.47 | +20 | +70 | −18 |
| L2 (001L_TL) | " | Tidally Locked | " | " | +28 | +95 | −18 |
| A1 (001M) | Aquaplanet | Present Day | " | " | +31 | +37 | −21 |
| A2 (001M2_TL) | " | Tidally Locked | " | " | +8 | +24 | −13 |
| V1 (001J) | Venus 310m | Present Day | " | " | +9 | +33 | −34 |
| V2 (001J_TL) | " | Tidally Locked | " | " | −4 | +49 | −44 |
| L1M (001Lc) | Venus 10m | Present Day | Modern | 1.9 | +26 | +71 | −4 |
| L2M (001Lc_TL) | " | Tidally Locked | " | " | +33 | +87 | −6 |
| A1M | Aquaplanet | Present Day | " | " | +38 | +43 | +32 |
| A2M | " | Tidally Locked | " | " | +16 | +37 | −1 |
| V1M (001N2) | Venus 310m | Present Day | " | " | +16 | +34 | −20 |
| V2M (001N2_TL) | " | Tidally Locked | " | " | +13 | +49 | −18 |

**Discussion**

For the first 6 entries in Table 1 Figure 1 shows images of the surface temperature averaged over 1/6 of a diurnal cycle for each non-tidally locked run and over 100 orbits for the tidally locked ones (there is too much variability on timescales comparable to those used for the non-TL runs). In Figure 2 we show the mean surface temperature versus two different insolations (what Venus received 2.9Gya and today) for each simulation in a combined plot. These demonstrate the importance of the expanded Hadley cell (due to slow rotation rates) and the subsequent large cloud cover at the substellar point and play the same role here as in the simulations of Paper I. With the exception of the high insolation land planet simulation (L2M) and the high insolation Aquaplanet (A1M) the planet is able to maintain liquid water across a large fraction of the surface and a dry stratosphere. For the two exceptions the situation is more complex. They are both warmer than most of the other simulations, and their stratospheres contain too much water (Fig 2B). L2M will lose what little water it has relatively quickly, while A1M will take perhaps a few billion years. At the same time several other of the high insolation runs are close to the limit and they may also lose their oceans over slightly longer timescales.

If one looks more closely at the temperature values presented it is clear that the land planets are the hottest of all of the simulations. The explanation for this is that in order to create the large convective cloud structure at the substellar point as described in Paper I it is necessary to have a lot of available moisture. In the land planet cases there is limited moisture available, hence it is not possible to form such a large contiguous cloud structure at the substellar point. This causes more radiation to reach the surface and raise the surface temperature relative to the other non-land planet simulations.

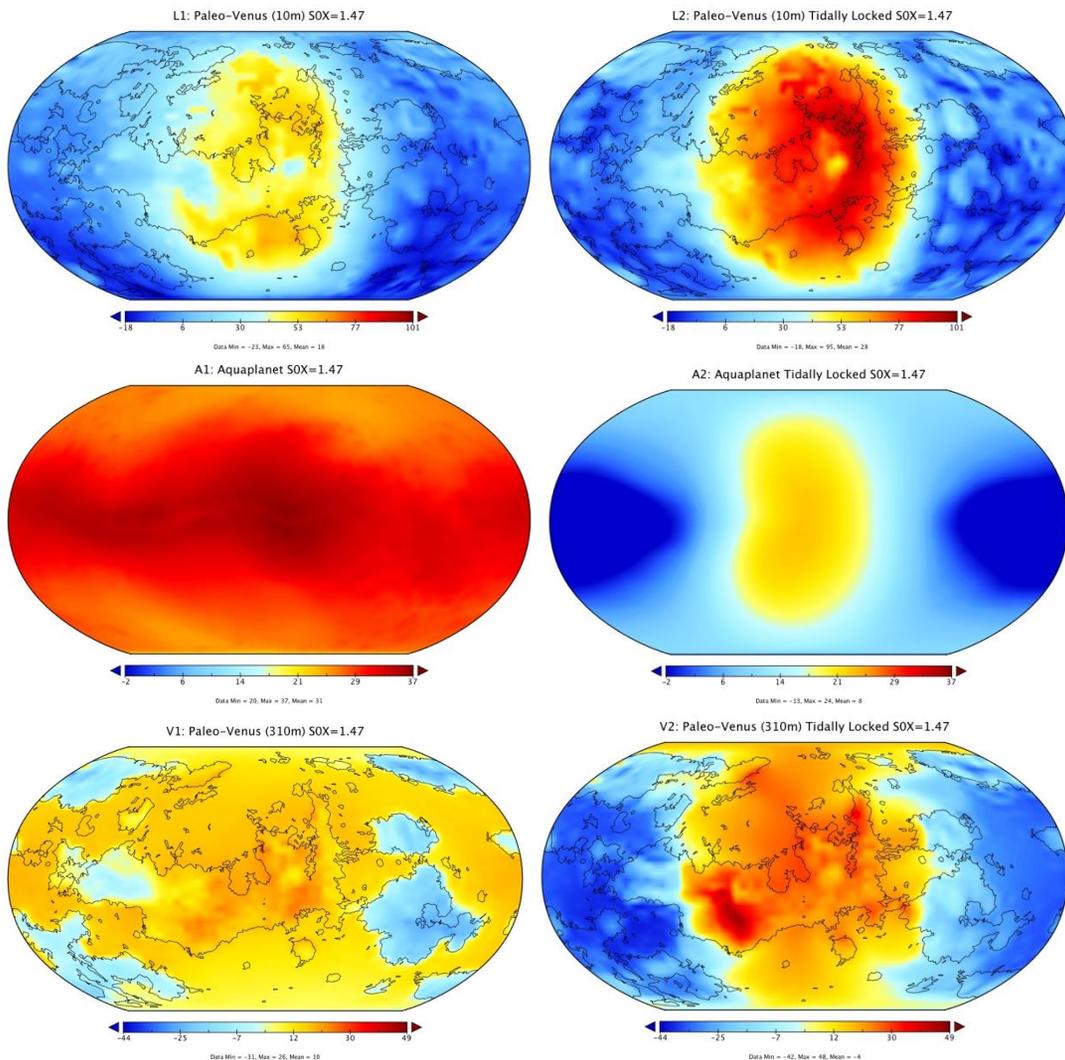

**Figure 1:** Surface temperature plots for the simulations listed in Table 1. Each set of 2 images have unique minimum and maximum scales. Non-tidally locked are averaged over 1/6[th] of a diurnal cycle, Tidally locked are averaged over 100 orbits. The substellar point is located at the center of each image.

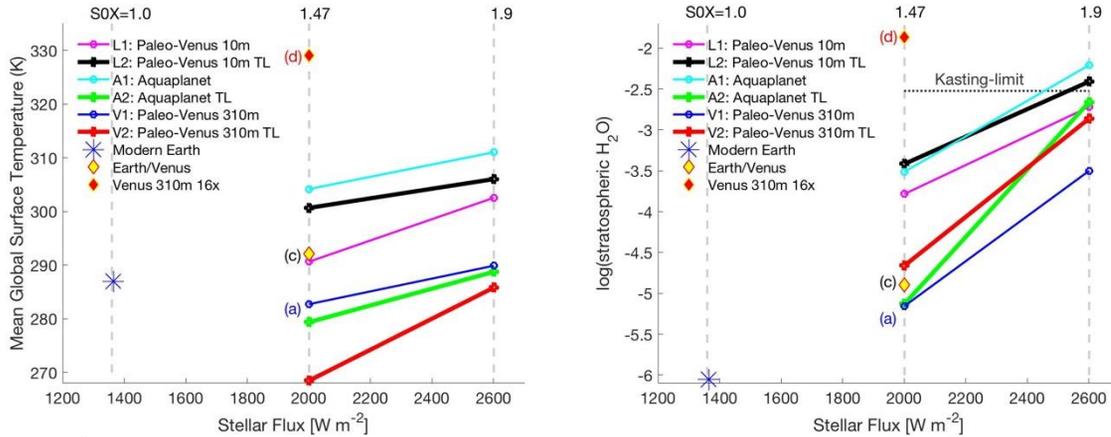

**Figure 2: (A) Left:** Plots of mean global surface temperature versus insolation for simulations listed in Table 1. Points identified with (a), (c), and (d) are the same as those in Table1/Figure 2 of Paper I. Earth/Venus refers to a simulation using Earth topography with a 310m deep bathtub ocean but with modern Venus orbital parameters and a solar spectrum from 2.9Gya (S0X=1.47, where S0X=1 is what Earth receives today as 1361W/m$^2$). Venus 310m 16x uses modern Venus topography and orbital parameters with a solar spectrum from 2.9Gya, but with a sidereal day length 16 times longer than that of modern day Earth. **(B) Right:** Similar to (A), but we plot log (stratospheric water vapor content) versus insolation. Values above -2.5 (labeled "Kasting-limit") signify that this world could lose 1 Earth ocean over a time period of ~4 billion years entering a moist greenhouse state (See Kasting 1988).

## Conclusions

Using a new suite of simulations we demonstrate that there are several ways in which Venus and Venus-like exoplanets could host liquid water on their surfaces for insolations from 1.47 to 1.9 times modern Earth's insolation. This may imply that the reason Venus lost its surface liquid water is not because of a moist greenhouse, or runaway greenhouse effect due to its close proximity to our sun as is commonly theorized. It could be that the resurfacing events over the last 1 billion years released too much CO$_2$ during that time, increasing the surface temperature and pushing the planet down the moist or runaway greenhouse pathway independent of the insolation received. In addition, these point to the fact that scientists should remain open minded when considering planets found in the Venus-Zone of nearby G-dwarf stars. They may prove more interesting than we ever imagined.

## Acknowledgements

The results reported herein benefited from participation in NASA's Nexus for Exoplanet System Science research coordination network sponsored by NASAs Science Mission Directorate. Resources supporting this work were provided by the NASA High-End Computing (HEC) Program through the NASA Center for Climate Simulation (NCCS) at Goddard Space Flight Center.

Thanks goes to Mareike Godolt for reminding MJW at a seminar in 2017 in Berlin that Jun Yang had previously explored insolations up to 1.9 times modern Earth in their 2013 study.

This research has made use of NASA's Astrophysics Data System Bibliographic Services.